\documentclass[pra,twocolumn,showpacs]{revtex4-1}
\usepackage{amsmath,amscd,amssymb,amsfonts}
\usepackage{graphicx,color}
\usepackage{epstopdf}
\usepackage{float}
\usepackage{enumerate}
\newcommand{\ie}{\textit{i.e.}}
\usepackage{xcolor}
\definecolor{maroon}{RGB}{100,20,20}
\definecolor{dblue}{RGB}{20,20,100}
\usepackage[colorlinks=true,linkcolor=dblue,citecolor=maroon,urlcolor=maroon]{hyperref}
\begin{document}
\title{Implementing efficient selective quantum process tomography of 
superconducting quantum gates on the IBM quantum processor}
\author{Akshay Gaikwad}
\email{ph16010@iisermohali.ac.in}
\affiliation{Department of Physical Sciences, Indian
Institute of Science Education \& 
Research Mohali, Sector 81 SAS Nagar, 
Manauli PO 140306 Punjab India.}
\author{Krishna Shende}
\email{ph19032@iisermohali.ac.in}
\affiliation{Department of Physical Sciences, Indian
Institute of Science Education \& 
Research Mohali, Sector 81 SAS Nagar, 
Manauli PO 140306 Punjab India.}
\author{Arvind}
\email{arvind@iisermohali.ac.in}
\affiliation{Department of Physical Sciences, Indian
Institute of Science Education \& 
Research Mohali, Sector 81 SAS Nagar, 
Manauli PO 140306 Punjab India.}
\author{Kavita Dorai}
\email{kavita@iisermohali.ac.in}
\affiliation{Department of Physical Sciences, Indian
Institute of Science Education \& 
Research Mohali, Sector 81 SAS Nagar, 
Manauli PO 140306 Punjab India.}
\begin{abstract}
The experimental implementation of selective quantum process
tomography (SQPT)  involves computing individual elements of
the process matrix with the help of a special set of states
called quantum 2-design states.  However, the number of
experimental settings required to prepare input states from
quantum 2-design states to selectively and precisely compute
a desired element of the process matrix is still high, and
hence constructing the corresponding unitary operations in
the lab is a daunting task.  In order to reduce the
experimental complexity, we mathematically reformulated the
standard SQPT problem, which we term the modified SQPT
(MSQPT) method.  We designed the generalized quantum circuit
to prepare the required set of input states and formulated
an efficient measurement strategy aimed at minimizing the
experimental cost of SQPT.  We experimentally demonstrated
the MSQPT protocol on the IBM QX2 cloud quantum processor
and selectively characterized various two- and three-qubit
quantum gates.
\end{abstract} 
\pacs{03.65.Wj, 03.67.Lx, 03.67.Pp, 03.67.-a} 
\maketitle
\section{Introduction}
\label{intro}
In the quest to build a real quantum computer, several
difficulties need to be overcome, which include pure state
initialization, implementing high fidelity quantum
operations, performing efficient and noise-free
measurements and protecting the quantum state against
decoherence.  Quantum state tomography
(QST)~\cite{leskowitz-pra-2004} and quantum process
tomography (QPT)~\cite{childs-pra-2001} are standard tools
that are extensively used for 
the characterization and  benchmarking of quantum information
processing devices and protocols.

Resource requirements for standard QST and QPT methods grow
exponentially with increasing system size, and hence several
novel methods have been designed that focus on simplifying
and reducing experimental complexity such as maximum
likelihood estimation~\cite{singh-pla-2016}, adaptive
quantum tomography~\cite{qi-quantum-inf-2017}, self-guided
tomography~\cite{Chapman-prl-2016}, ancilla-assisted
tomography~\cite{altepeter-prl-2003}, compressed sensing
tomography~\cite{badveli-pra-2020,gianani-prxquantum-2020},
and least square optimization based
tomography~\cite{gaikwad-ijqi-2020,gaikwad-qip-2021}.  In
some cases, instead of the complete characterization of the
quantum process, one is only interested in a specific part,
and the method used is termed selective quantum process
tomography (SQPT)~\cite{paz-pra-2009,perito-pra-2018}.

These novel QST and QPT protocols have been experimentally
demonstrated on various physical configurations such as
NMR~\cite{wu-jcp-2013,gaikwad-pra-2018,glaser-pra-2018},
linear-optics~\cite{paz-prl-2011,teo-pra-2020},
NV-centers~\cite{zhang-prl-2014}, ion-trap based quantum
processors~\cite{riebe-prl-2006}, photonic
qubits~\cite{kim-prl-2020}, and superconducting
qubits~\cite{chow-prl-2009,kofman-pra-2009,rodionov-prb-2014,von-prxquantum-2020}.
It has been shown that sequential weak value measurement can
be used to perform direct QPT of a qubit
channel~\cite{zhang-annal-2017,kim-natcomm-2018}.  
A unitary 2-design and a twirling QPT protocol have been
used to certify a seven-qubit entangling gate on an NMR
quantum processor~\cite{lu-prl-2015}.
In recent
years, researchers across the globe are engaged in building
quantum systems of a larger register size termed noisy
intermediate-scale quantum (NISQ) processors, such as the
IBM quantum processor based on superconducting technology
with 32 qubits, and NMR, ion-trap based quantum computers
and linear optical photonic quantum processors which have
achieved register sizes of 12, 10 and 14 qubits,
respectively~\cite{negrevergne-prl-2006,gao-nature-phyics-2010,monz-prl-2011}.

In this work we demonstrate a modified SQPT (MSQPT) protocol
on a five-qubit IBM QX2 quantum information processor and
use it to characterize several two- and three-qubit
superconducting quantum gates.  We propose a general quantum
circuit for initial input state preparation to efficiently
implement the MSQPT protocol.  We implement an efficient
measurement framework wherein detection is performed on only
a single qubit.  Our experimental results show that one can
use the modified SQPT protocol to efficiently and
selectively characterize the desired quantum process.  We
demonstrate that the MSQPT results can be further refined to
construct the underlying true quantum process by solving a
constrained convex optimization problem.

This paper is organized as follows:~
The
modified SQPT protocol designed for implementation on the
IBM quantum processor is described in Section~\ref{sec2}.
The  mathematical
formulation of standard SQPT is given in Section~\ref{sec2.1}.  
The MSQPT protocol and quantum circuit are described in 
Section~\ref{sec2.2}.
The steps to construct the complete
set of initial input quantum states for arbitrary dimensions
and the construction of the
unitary are described in Section~\ref{sec2.3}.  
Sections~\ref{2qubit} and \ref{3qubit} contain details of the
experimental implementation of the MSQPT method for two and
three qubits on the IBM quantum processor, respectively.
Section~\ref{concl} presents some concluding remarks and
future directions.
\section{Modified protocol for selective quantum process
tomography for the IBM quantum computer}
\label{sec2}
\subsection{Standard selective quantum process tomography}
\label{sec2.1}
A quantum process denoted by the superoperator $\Lambda$
can be described using the Kraus operator 
representation~\cite{kraus-book-1983}:
\begin{equation} 
\Lambda(\rho)=\sum_{m,n} \chi_{mn} {E_m} \rho
{E_n}^\dagger. 
\label{eq2} 
\end{equation}
with $\lbrace E_i \rbrace$ being a  fixed set of basis
operators, and 
$\rho$ being the quantum state evolving under
$\Lambda$. The matrix $\chi$ with elements $\chi_{mn}$ 
characterizes the given quantum process
$\Lambda$. Estimating the complete matrix $\chi$ is referred
to as performing QPT of $\Lambda$.  Full QPT is achieved by
preparing a complete set of linearly independent quantum
states $\lbrace \rho_i \rbrace $ and then letting them
evolve under the  quantum process under
consideration~\cite{chuang-jmo-09}.
However, sometimes it suffices to estimate specific
elements of the $\chi$ matrix, a procedure referred to as
selective QPT (SQPT), with an experimental complexity which
is lower than the full QPT
protocol~\cite{paz-pra-2009}.

A specific element $\chi_{mn}$ of the process 
matrix can be determined by 
computing `average survival probabilities' $F_{mn}$ 
as~\cite{paz-prl-2011}: 
\begin{equation}
F_{mn} = \frac{1}{K}
\sum_{j} \langle \phi_j \vert \Lambda (E_m^\dagger \vert
\phi_j \rangle \langle \phi_j \vert E_n)\vert \phi_j \rangle
=\frac{ D\chi_{mn}+\delta_{mn}}{D+1} 
\label{eq3} 
\end{equation} 
where $ \lbrace \vert \phi_i \rangle \rbrace $ are a set of
quantum 2-design states~\cite{gaikwad-pra-2018}, $K$ is
their cardinality and $D$ is the dimension of the Hilbert
space.  
We rewrite
the operators $(E_m^\dagger
\vert \phi_j \rangle \langle \phi_j \vert E_n)$ and
$\Phi_j=\vert \phi_j \rangle \langle \phi_j \vert$ 
in Eq.~\ref{eq3}
in terms
of fixed basis operators $\lbrace E_i \rbrace$ as
$E_m^\dagger \Phi_j E_n= \sum_{i} {^jc}_i^{mn}E_i $ and
$\Phi_j=\sum_{k} {^je_k E_k}$ (with $^je_k\in\mathbb{R}$),
which leads to the compact form:
\begin{equation}
F_{mn}= \frac{1}{K} \sum_{i,j,k} {^j\beta^{mn}_{ki}} \textbf{Tr}
\left[E_k \Lambda(E_i)\right]
\label{eq5}
\end{equation}
where the complex scalar quantities ${^j\beta^{mn}_{ki}}={}^je_k
\,{}^jc^{mn}_i$ can be computed analytically and do not
depend upon the quantum process. It turns out that if we
choose Pauli matrices as basis operators, 
then for given values of $m$ and $n$, the tensor
${^j\beta^{mn}_{ki}}$ is sufficiently sparse and most of its
values are zero. We hence only need to compute $\textbf{Tr}[E_k
\Lambda(E_i)] \equiv \bar{E}_k^i$ for those  values
of $i$ and $k$ for which ${^j\beta^{mn}_{ki}} \neq 0$.
The sparsity of ${^j\beta^{mn}_{ki}}$ is directly connected to
the experimental complexity in terms of the number of
coefficients $\bar{E}_k^i$ that need to be estimated.

The question now arises
about the estimation of the coefficients $\bar{E}_k^i$.
Given a set of operators $E_i$($n$-qubit Pauli operators), 
one can associate a
well defined (positive and unit trace) density operator 
$\tilde{\rho_i}$ with it 
as follows:
\begin{eqnarray} 
	\tilde{\rho_0} &=&\frac{1}{D}E_0\nonumber \\
	\tilde{\rho}_i &=& \frac{1}{D}(E_i + I), \quad i>0
\label{eq8}
\end{eqnarray}
It is easy to see that 
\begin{equation}
	\bar{E}_k^i = \textbf{Tr}[E_k
\Lambda(E_i)] = D \textbf{Tr}[E_k \Lambda(\tilde{\rho_i})] 
\label{rho-construct}
\end{equation}
Eq.~\ref{rho-construct} hinges on the fact
that the identity operator does not evolve under the process
matrix $\Lambda$. This provides us with a way to
experimentally estimate the desired coefficients
$\bar{E}_k^i$, where we need to prepare the system in states
$\tilde{\rho}_i$, let it evolve under the process $\Lambda$
and then measure $E_k$.
\subsection{Protocol for MSQPT}
\label{sec2.2}
Although the SQPT protocol is computationally less
resource-intensive as compared to the standard QPT method,
the number of experimental settings required to prepare the
input states for computing a selected element of the process
matrix is still quite high. We propose a generalization of
the SQPT method, namely the MSQPT
protocol, which considerably reduces the experimental
complexity of computing a desired element of the process
matrix with high precision. The computational efficiency
of the MSQPT protocol is based on the fact that the total
number of input states that are required to calculate the
average survival probabilities (Eq.~\ref{eq3}) is much fewer
as compared to the SQPT method, as a single unitary operator
is applied simultaneously on all system qubits to prepare
the input state.  Furthermore,  only a single detection is
required at a time, which reduces the number
of readouts required to determine a specific element of the
process matrix, further reducing the experimental complexity
of the protocol.

\begin{figure}[ht]
\includegraphics[angle=0,scale=1]{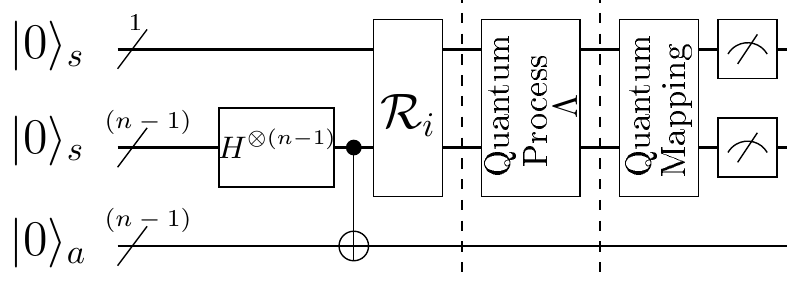} 
\caption{
The quantum circuit to acquire data to perform an 
$n$-qubit MSQPT. The symbol '/' through the input wire
represents a multiqubit quantum register.  The first and the
second quantum registers collectively represent the system
qubits (denoted by $\vert 0 \rangle_s$), and the third
quantum register represents the ancilla qubits (denoted by
$\vert 0 \rangle_a$).  The first block prepares the desired
pure input state $\vert\Psi_i\rangle$.  The unitary gate
$\mathcal{R}_i$ is then applied on the system qubits.  The
second block represents the unknown quantum process which is
to be applied to the system qubits and the last block
represents the measurement settings to compute the
expectation value of the desired observable. 
} 
\label{genckt} 
\end{figure}

The quantum circuit to implement the $n$-qubit
MSQPT protocol is given in Fig.~\ref{genckt}.  The symbol
'/' through the input wire represents a multiqubit quantum
register.  The first quantum register contains a single
qubit while the second and third quantum registers comprise
$n-1$ qubits, respectively.  The first and the second quantum registers
collectively represent the system qubits denoted by $\vert 0
\rangle_s$, while the third quantum register represents the
ancilla qubits denoted by $\vert 0 \rangle_a$.  The first
block prepares the desired pure input state
$\vert\Psi_i\rangle$, where
$H^{\otimes (n-1)}$ is applied on the second register
followed by $n-1$ CNOT gates, with the control being  at the
second quantum register and the target being at the third
quantum register.  The unitary gate
$\mathcal{R}_i$ is then applied on the system qubits, where the
columns of the unitary operation $\mathcal{R}_i$ are the
normalized eigenvectors of the density matrix $\tilde{\rho}_i$.
The second block represents the unknown quantum process
which is to be characterized and the last block represents
the measurement settings to compute the expectation values
of the desired observables. Note that in the third block,
after the appropriate quantum mapping, only a single
detection is performed at a time, to measure a desired
observable.

In order to represent a valid quantum map, the
$\chi$ matrix should satisfy following
conditions\cite{obrien-prl-04}: (i) $\chi = \chi^{\dagger}$,
(ii) $\chi \geq 0$ and (iii)
$\sum_{m,n}\chi_{mn}E_m^{\dagger}E_n = I$. Using the MSQPT
method, the $\chi$ matrix is Hermitian by construction,
however there
is no guarantee that it will satisfy the last two conditions.
One can use the constrained convex optimization (CCO)
technique~\cite{gaikwad-qip-2021} to obtain a valid
$\chi_{\rm cco}$ matrix from $\chi_{\rm msqpt}$ as follows:
\begin{subequations}
\begin{alignat}{2}
&\!\min_{\chi}      &\qquad& \Vert \chi_{\rm
msqpt}-\chi_{\rm cco}\Vert_{l_2}\label{e12}\\
&\text{subject to} &    & \chi_{\rm cco} \geq 0,\label{e12:constraint1}\\
&   &    & \sum_{m,n}\chi_{mn}^{\rm
cco}E_m^{\dagger}E_n = I.\label{e12:constraint2}
\end{alignat}
\end{subequations}
where $\chi_{\rm msqpt}$ is the experimentally obtained process
matrix using the MSQPT protocol and $\chi_{\rm cco}$ is 
the variable
process matrix which represents the underlying true quantum
process.

\subsection{State preparation and unitary operator
construction}
\label{sec2.3}
We note here that for an $n$-qubit system, all density operators
$\tilde{\rho_i}$ in Eq.~\ref{eq8} represent mixed states (except
for $n=1$).  We hence require ancillary qubits to experimentally
prepare the quantum system in the state $\tilde{\rho_i}$. 

It turns out that for an $n$-qubit
system, all non-zero eigenvalues of the operator
$\tilde{\rho_i}$ in
Eq.~\ref{eq8} are the same and are equal to $1/2^{n-1}$. Let
$\lbrace \vert u_1^i \rangle, \vert u_2^i \rangle, \vert
u_3^i \rangle,...,\vert u_{2^{n-1}}^i \rangle \rbrace $
represent the
complete set of normalized eigenvectors of the operator
$\tilde{\rho_i}$ corresponding to its non-zero eigenvalues. The state
of the combined system (system $+$ ancilla) we need to
prepare
is given by:
\begin{equation}\label{geneq}
\vert \Psi_i \rangle = \frac{ \vert u_1^i \rangle \vert
a_1\rangle+ \vert u_2^i \rangle \vert a_2\rangle+.....+
\vert u_{2^{n-1}}^i \rangle \vert
a_{2^{n-1}}\rangle}{\sqrt{2^{n-1}}}
\end{equation}
where $\vert a_i\rangle$ are the basis states 
of the ancilla qubits. Note that in general
$\vert\Psi_i\rangle$ represents an entangled state. After
tracing over the ancillary qubits, the system
will be in the desired state $\tilde{\rho_i}$.

The unitary operator $U^i$, such that $U^i \vert 0
\rangle ^{sys} \vert 0 \rangle ^{ancilla} = \vert \Psi_i
\rangle $ can be constructed as follows:
\begin{enumerate}
\item Apply a Hadamard gate on ($n-1$) 
system qubits; $2^{n-1}$ number of states will be in
a superposition state while the ancilla qubits will be
in the state $\vert 0 \rangle ^{ancilla}$.
\item Apply CNOT gates with the
system qubits being the controls and 
ancilla qubits being the target. We hence have $\vert 0
\rangle ^{ancilla} \longrightarrow \vert a_1^i \rangle $,
$\vert 0 \rangle ^{ancilla} \longrightarrow \vert a_2^i
\rangle $, and so on.
\item Map the computational basis
states of the system qubits 
to the eigenvectors of $\tilde{\rho_i}$ using the unitary gate
$\mathcal{R}_i$, where the columns of $\mathcal{R}_i$ are
the normalized eigenvectors of $\tilde{\rho_i}$ (Eq.~\ref{eq8}). Note that the
column position of eigenvectors depends on which
computational basis vector we want to map onto which
eigenvector. 
The combined system (system $+$ ancilla qubits)
will be in the $\vert\Psi_i\rangle$ state.
\item 
Repeat the steps [1-3] to prepare other states $\tilde{\rho_i}$.
\end{enumerate}

\begin{figure*}[ht]
\includegraphics[angle=0,scale=1]{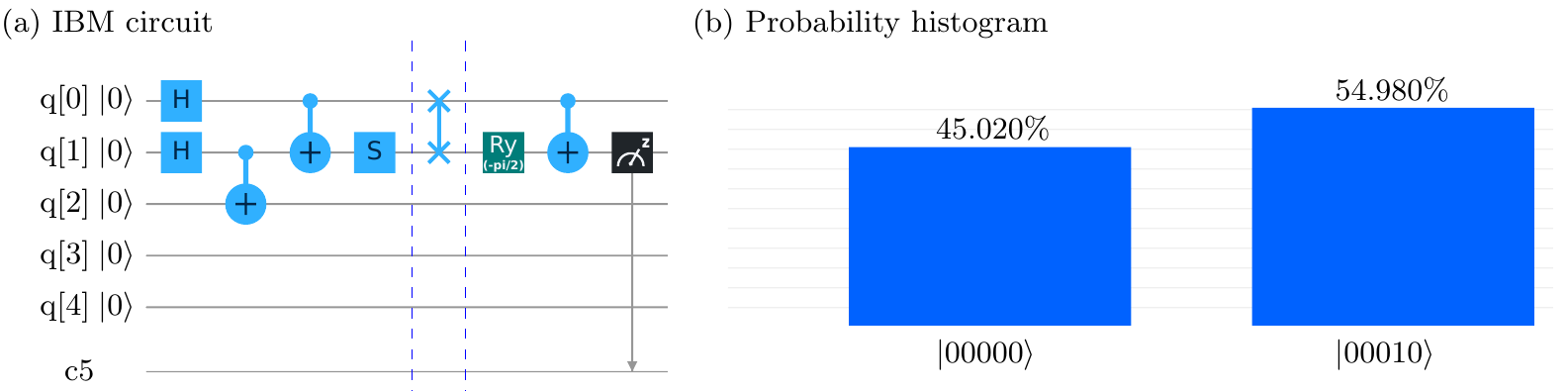} \caption{(a)
The IBM quantum circuit to perform MSQPT of a two-qubit SWAP
gate.  The first block prepares the three-qubit input state
$\vert \psi_{6} \rangle$.  The quantum process corresponding
to the two-qubit SWAP gate is applied in the second block
and in the last block, the quantum map
$U_{13}=$CNOT$_{12}$. $R_y(-\frac{\pi}{2})$ is applied to
compute ${\rm Tr}(\sigma_z \otimes \sigma_x
\Lambda(\tilde{\rho_6}))$ by detecting the second qubit in the
$\sigma_z$ basis. (b) The histogram representing statistical
results after implementing the quantum circuit given in (a),
$4096$ times. The values $p_0 = 0.4502$ and $p_1 = 0.5498$
represent the probabilities of obtaining the second qubit in
the $\vert 0 \rangle $ and the $\vert 1 \rangle $ state,
respectively.} 
\label{ibm} 
\end{figure*}
\section{Experimental implementation of MSQPT on the IBM
QX2 processor}
\label{sec4}
The IBM quantum processor is based on superconducting qubits
and is freely available through the
cloud~\cite{santos-2017,harper-prl-2019,Rui2017}, and has
been used to demonstrate various quantum
protocols~\cite{behera-qip-2017,joy-qip-2019}.  
More details about the architecture of the
IBM QX2 processor and the topology of superconducting qubits
are given in \cite{devitt-pra-2016} and information
about the form of the Hamiltonian and important relaxation
parameters can be found in~\cite{pathak-pla-2020,dueck-conf-2018}.  
We use the
five-qubit IBM QX2 processor to demonstrate the
MSQPT protocol described in the previous section. The
system is prepared in an input state corresponding to all
qubits being in the $\vert 0 \rangle $ state. 
After the
gate implementation, projective measurements are performed in the
Pauli $\sigma_z$ basis and the quantum circuit is
implemented multiple times to compute the Born
probabilities.  
The IBM
quantum architecture requires a pure quantum state as an
input state and only allows the implementation of unitary
operations. We hence utilize ancillary qubits to prepare
the system in a mixed state and to simulate non-unitary
evolution.

We implement the MSQPT protocol
corresponding to two- and three-qubit gates
and elementwise construct the corresponding full $\chi$
matrices. In all the cases 
considered, we use the experimentally constructed
$\chi_{\rm msqpt}$, solve the CCO problem 
(Eq.~\ref{e12}) and obtain $\chi_{\rm cco}$, which represents
the underlying true quantum process. 
The
fidelity of experimentally implemented quantum gates is
computed using the measure~\cite{zhang-prl-2014}:
\begin{equation}
{\mathcal F}(\chi^{}_{{\rm exp}},\chi^{}_{{\rm the}})=
\frac{|{\rm Tr}[\chi^{}_{{\rm exp}}\chi_{{\rm the}}^\dagger]|}
{\sqrt{{\rm Tr}[\chi_{{\rm exp}}^\dagger\chi^{}_{{\rm exp}}]
{\rm Tr}[\chi_{{\rm the}}^\dagger\chi^{}_{{\rm the}}]}}
\label{fid}
\end{equation} 
To validate
our circuits, we also theoretically simulate the MSQPT
protocol on the IBM processor and obtain $\chi_{{\rm sim}}$. 
The fidelity of the simulated quantum gates is computed by
using a similar measure as given in Eq.~\ref{fid}.

\subsection{MSQPT of two-qubit quantum gates}
\label{2qubit}
For two qubits, we need to prepare
fifteen input (mixed) states  $\tilde{\rho}_i$ (Eq.~\ref{eq8})
corresponding to all the Pauli operators $E_i$. For all
$\tilde{\rho}_i$s, it
turns out that out of four eigenvalues, only two eigenvalues
are non-zero ($\lambda_1 = \lambda_2 = 1/2$).
Let $\vert
v^i_1 \rangle$ and $\vert v^i_2 \rangle$ represent the
normalized eigenvectors of the operator $\tilde{\rho_i}$
corresponding to $\lambda_1$ and $\lambda_2$, respectively.
To perform MSQPT of two qubits  on the IBM computer, we
use one ancillary qubit and prepare
three-qubit input (pure) states of the form:
\begin{equation}
\vert \psi_i \rangle = \frac{\vert v^i_1 \rangle \vert 0 \rangle + \vert v^i_2 \rangle \vert 1 \rangle }{\sqrt{2}}
\end{equation}

All fifteen three-qubit pure input states $\vert \psi_i \rangle$
corresponding to $E_i$ are listed below:

\begin{align*}
& \vert \psi_1 \rangle =[(0,1,0,1,1,0,1,0)/2]^{T},  \\&
\vert \psi_2 \rangle=[(0,-i,0,1,-i,0,1,0)/2]^T,  \\&
\vert \psi_3 \rangle=[(1,1,0,0,0,0,0,0)/\sqrt{2}]^T, \\&
\vert \psi_4 \rangle=[(0,-1,1,0,0,-1,1,0)/2]^T, \\&
\vert \psi_5 \rangle=[(1,0,0,1,0,1,1,0)/2]^T, \\&
\vert \psi_6 \rangle=[(-i,0,0,1,0,-i,1,0)/2]^T, \\&
\vert \psi_7 \rangle=[(0,1,-1,0,0,1,1,0)/2]^T, \\&
\vert \psi_8 \rangle=[(0,-1,-i,0,0,-i,1,0)/2]^T, \\&
\vert \psi_9 \rangle=[(-i,0,0,1,0,i,1,0)/2]^T, \\&
\vert \psi_{10} \rangle=[(-1,0,0,1,0,1,1,0)/2]^T, \\&
\vert \psi_{11} \rangle=[(0,1,1,0,0,1,1,0)/2]^T, \\&
\vert \psi_{12} \rangle=[(1,0,0,1,0,0,0,0)/\sqrt{2}]^T, \\&
\vert \psi_{13} \rangle=[(0,1,0,1,-1,0,1,0)/2]^T, \\&
\vert \psi_{14} \rangle=[(0,-i,0,1,i,0,1,0)/2]^T, \\&
\vert \psi_{15} \rangle=[(1,0,0,0,0,0,0,1)/\sqrt{2}]^T
\end{align*}

As an illustration, the IBM quantum circuit for 
implementing MSQPT of a two-qubit SWAP gate,
corresponding to the quantum state $\vert \psi_6 \rangle$ and
the observable $E_{13} = \sigma_z \otimes \sigma_x$, is given in
Fig.~\ref{ibm}. 
The system qubits are denoted by
$q[0]$ and $q[1]$ (the first and second qubit, respectively)
while the ancilla qubit is denoted by $q[2]$. 
To prepare the system in the pure state $\vert \psi_6
\rangle$, the unitary operation $U^6 = S_2.$
CNOT$_{12}. $CNOT$_{23}.$ H$_2.$ H$_1$ is applied on the
initial state $\vert 000 \rangle$ in the first block. 
In the
second block, the quantum process $(\Lambda_{system} \otimes
I_{ancilla})$ corresponding to a two-qubit SWAP gate is
implemented on the system qubits. 
In the last block, the quantum
map corresponding to the unitary operation $U_{13}=$
CNOT$_{12}.R_y(-\frac{\pi}{2})$ is used to transform the
output state and determine $E_{13}=\langle \sigma_z \otimes
\sigma_x \rangle $ by measuring the second qubit in the $
\sigma_{z} $ basis~\cite{gaikwad-pra-2018}. 
The quantity
corresponding to ${\rm Tr}(\sigma_z \otimes \sigma_x
\Lambda(\tilde{\rho_6}))$ is experimentally computed, which is 
equivalent to
${\rm Tr}(\sigma_{2z}
U_{13}(\Lambda(\tilde{\rho_6})){U_{13}}^{\dagger})$.
Using
Eq.~\ref{eq8} we obtain:
\begin{eqnarray}
{\rm Tr}(\sigma_z \otimes \sigma_x \Lambda(\sigma_x \otimes \sigma_y )) 
&=& 4 {\rm Tr}(\sigma_z \otimes \sigma_x
\Lambda(\tilde{\rho_6})) \nonumber \\
 &=& 4 {\rm Tr}(\sigma_{2z} U_{13}(\Lambda(\tilde{\rho_6})){U_{13}}^{\dagger})
\nonumber \\
\end{eqnarray}
One can thus efficiently compute all the $ \langle E_k^i
\rangle$ (Eq.~\ref{eq5}) and estimate the corresponding
average survival probabilities $F_{mn}$.
The list of all unitary
operations $U_i$ corresponding to all quantum maps which
transform output states in order to determine $\langle E_k
\rangle $ by detecting either of the system qubits in
the $\sigma_z$ basis (\ie by measuring either $\langle
\sigma_{1z} \rangle$ or $\langle \sigma_{2z} \rangle$) is
given in \cite{gaikwad-pra-2018}.
\begin{figure*}[ht]
\includegraphics[angle=0,scale=1]{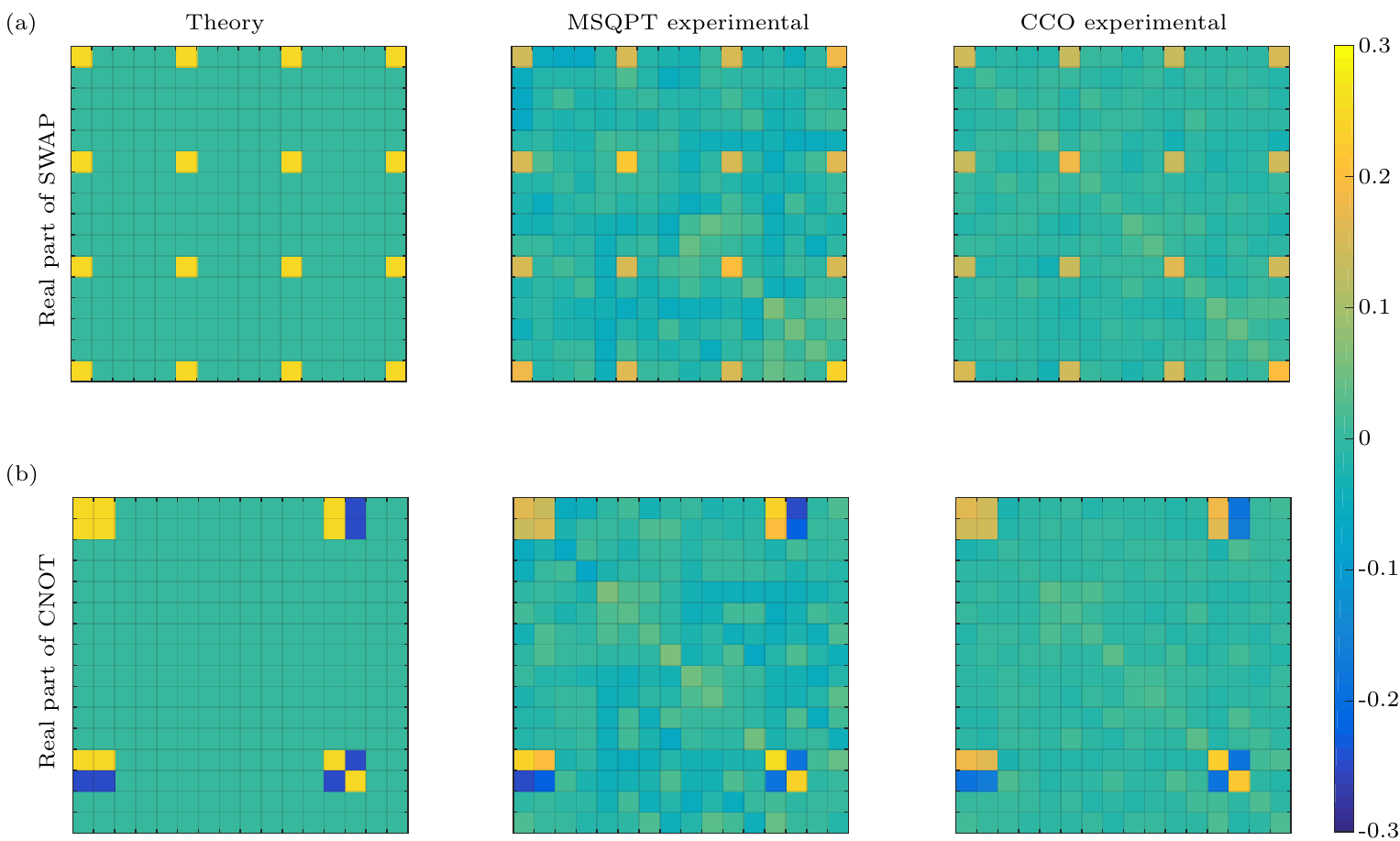} 
\caption{Matrix plots corresponding to the real part of the
(a) $\chi$ matrix for the SWAP gate and (b) for the CNOT
gate. The first column represents the theoretically constructed
process matrix $\chi_{{\rm the}}$, while the second and third
columns represent $\chi_{{\rm msqpt}}$ and $\chi_{{\rm cco}}$,
respectively.}
\label{plot1}
\end{figure*}
\begin{table}[ht]
\caption{\label{complexity}
Experimental complexity and the number of ancillary qubits
required for the implementation of two-qubit MSQPT, SQPT 
and standard QPT protocols.}
\begin{ruledtabular}
\begin{tabular}{c c c c}
&
MSQPT&
SQPT&
Standard QPT~~~\\
\colrule
Preparations & 15 & 80 & 15 ~~~\\
Readouts & 60 & 240 & 225 ~~~\\ 
Ancilla qubits & 1 & 0 & 0 ~~~\\   
\end{tabular}
\end{ruledtabular}
\end{table}

The $16\times 16$ grid matrix plots in Fig.~\ref{plot1}(a)
represent the real part of the $\chi$ matrix corresponding
to the two-qubit SWAP gate, where the position of the specific grid
represents the corresponding element of the $\chi$ matrix,
while its color represents its value.  For instance, the
first yellow square in the matrix plot in
Fig.~\ref{plot1}(a) denotes the element $\chi_{11}=0.25$ of
the theoretically constructed process matrix $\chi_{\rm
the}$.  
Only 16 yellow squares 
have non-zero values  
in the theoretically constructed
matrix plot for the SWAP gate. 
The second and third
columns  represent matrix plots corresponding $\chi_{\rm
msqpt}$, and $\chi_{\rm cco}$ respectively obtained by
implementing MSQPT protocol on IBM QX2 processor.  The
differences in the theoretically computed and experimentally
obtained matrix plots reflect errors due to decoherence and
statistical and systematic errors while preparing the
initial input state.  The matrix plots corresponding to the
imaginary part of $\chi$ matrix are not presented, since all
the elements of imaginary part of the $\chi$ matrix are
zero.  The color grids in the matrix plots  in
Fig.~\ref{plot1} in the third column (CCO experimental) have
a smaller deviation as compared to the matrix plots in the
second column (MSQPT experimental).  This improved fidelity
implies that one can use the MSQPT data to solve CCO problem
and reconstruct the full process matrix more accurately.
The experimental fidelity of $\chi_{{\rm
msqpt}}$ for the SWAP gate (Fig.~\ref{plot1}(a)) turned out
to be 0.799, while the improved fidelity of $\chi_{{\rm
cco}}$ turned out to be 0.929.  
We also computed the process matrices for the two-qubit
CNOT gate and the corresponding matrix plots are
shown in Fig.~\ref{plot1}(b).
The experimental fidelity of $\chi_{{\rm msqpt}}$ for the
CNOT gate turned out to be 0.828,
while the improved fidelity of $\chi_{{\rm cco}}$ turned out
to be 0.953.  
We obtained fidelities of
$\mathcal {F}(\chi_{{\rm sim}})\geq 0.99$ for all the
quantum gates, which ensures that all the quantum circuits
are correct.  The fidelity values of $\mathcal
{F}(\chi_{{\rm cco}})\geq 0.9$ shows that one can retrieve
the full dynamics of the quantum process with considerably
high precision by solving the optimization problem
(Eq.~\ref{e12}) using the experimentally constructed full
$\chi_{{\rm msqpt}}$ matrix.

The standard QPT protocol is based on the linear inversion
method and requires the preparation of 15 linearly
independent input states and further requires the state
tomography of each output state.  Hence the total number of
readouts to determine a specific element of the 
two-qubit process matrix with high precision, using the
standard QPT protocol, is  $15 \times 15 = 225$.  The SQPT
protocol uses quantum two-design states as initial input
states and further requires a quantum operation to prepare
the system in the desired state. Determining the real and
imaginary parts of $F_{mn}$ respectively requires a total of
80 state preparations.  Further, to estimate the overlap with
original state $\vert \phi_j \rangle \langle \phi_j \vert$
(Eq.~\ref{eq3}), three readouts need to be performed
(as there are three non-zero coefficients in the
decomposition of $\Phi_j$).  Hence the total number of
readouts to determine a specific element of  the 
two-qubit process matrix  with high precision, using the
SQPT protocol, is  $80 \times 3 =240$.  In the MSQPT protocol,
the average survival probabilities can be computed quite
efficiently as the total number of states that need to be
prepared are only 15, there are only 12 readouts per 
mutually unbiased basis (MUB)
set, and there are 5 MUB sets which form a complete set of
quantum 2-design states.  Hence the total number of
readouts to determine a specific
element of the two-qubit process matrix with high
precision, using the
MSQPT protocol, is  only $12 \times 5 = 60$.  
The experimental
complexity and number of ancilla qubits required to
determine a specific element of the process matrix 
with high precision for a two-qubit system, using  the MSQPT method, are
compared with the standard QPT and SQPT methods in
Table~\ref{complexity}.

\begin{figure*}[ht]
\includegraphics[angle=0,scale=1]{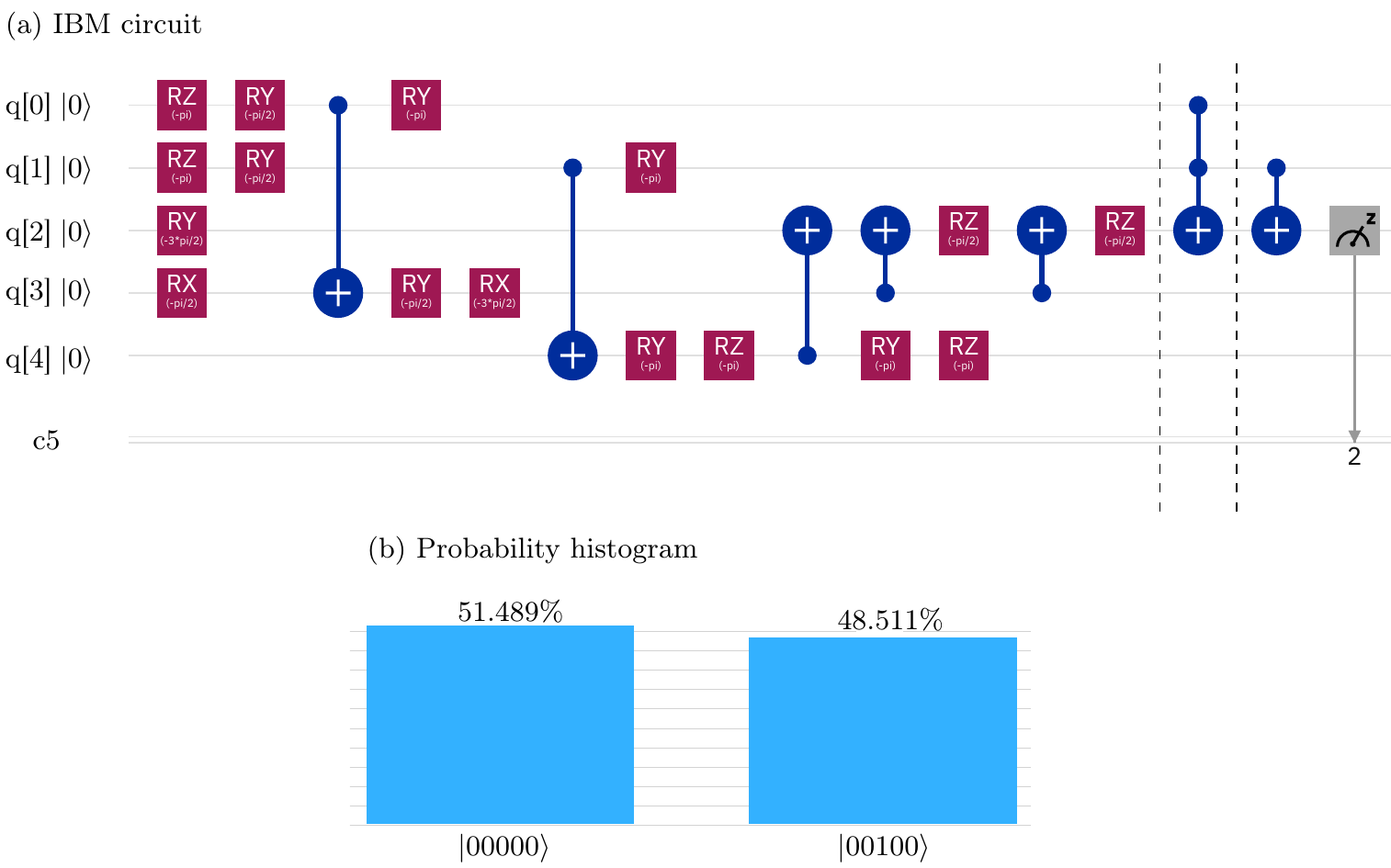} 
\caption{(a) The IBM quantum circuit to perform MSQPT of 
a three-qubit Toffoli gate. 
The first block prepares the five-qubit input 
state $\vert \Omega_{50} \rangle $.
The  quantum process corresponding to
the Toffoli gate is applied in the
second block, and
in the last block, the quantum map $U_{15}= $CNOT$_{23}$ 
is applied to compute
${\rm Tr}(I \otimes \sigma_z \otimes \sigma_y
\Lambda(\tilde{\rho}_{50}))$ by detecting the third qubit
in the $\sigma_z$ basis. (b) The histogram represents statistical
results after implementing the quantum circuit given in (a), 4096
times. The values $p_0 = 0.51489$ and $p_1 =
0.48511$ represent the probabilities of obtaining 
the third qubit in the $\vert 0 \rangle $ and
the $\vert 1 \rangle $ state, respectively.} 
\label{ibm4}
\end{figure*}

\subsection{MSQPT of three-qubit quantum gates}
\label{3qubit}
To perform MSQPT on a three-qubit system, we need to prepare 63
input (mixed) states $\tilde{\rho_i}$ corresponding to all
the three-qubit Pauli operators $E_i$. It turns out that for all
$\tilde{\rho_i}$, out of 8 eigenvalues only 4 are non zero and
are equal to 1/4. Let $\vert u_1^i \rangle$, $ \vert u_2^i
\rangle$, $ \vert u_3^i \rangle$ and $\vert u_4^i \rangle$
be the 4 eigenvectors of $\tilde{\rho_i}$ with
non-zero eigenvalues. In order to prepare the system in the
any of the $\tilde{\rho_i}$ states, we 
first need to prepare a five-qubit pure state:
\begin{equation}
\vert \Omega_i \rangle = \frac{\vert u^i_1 \rangle \vert 00 \rangle + \vert u^i_2 \rangle \vert 01 \rangle + \vert u^i_3 \rangle \vert 10 \rangle + \vert u^i_4 \rangle \vert 11 \rangle }{2}
\label{eq14}
\end{equation}
After tracing out the two ancilla qubits, 
the three system qubits are in the state $\tilde{\rho_i}$, \ie,
${\rm Tr}^{ancilla}(\vert \Omega_i \rangle \langle
\Omega_i \vert) = \tilde{\rho_i}$. The list of all five-qubit pure
input states $\lbrace \vert \Omega_i \rangle \rbrace $ is
given in Appendix~\ref{sec.appendix1}.

\begin{table}[h!]
\caption{\label{complexity3}
Experimental complexity and the number of ancilla qubits
required for the implementation of three-qubit MSQPT, SQPT and
standard QPT protocols.}
\begin{ruledtabular}
\begin{tabular}{c c c c}
&
MSQPT&
SQPT&
Standard QPT~~~\\
\colrule
Preparations & 63 & 288 & 63 ~~~\\
Readouts & 504 & 2016 & 3969 ~~~\\ 
Ancilla qubits & 2 & 0 & 0 ~~~\\   
\end{tabular}
\end{ruledtabular}
\end{table}
\begin{figure*}[ht]
\includegraphics[angle=0,scale=1]{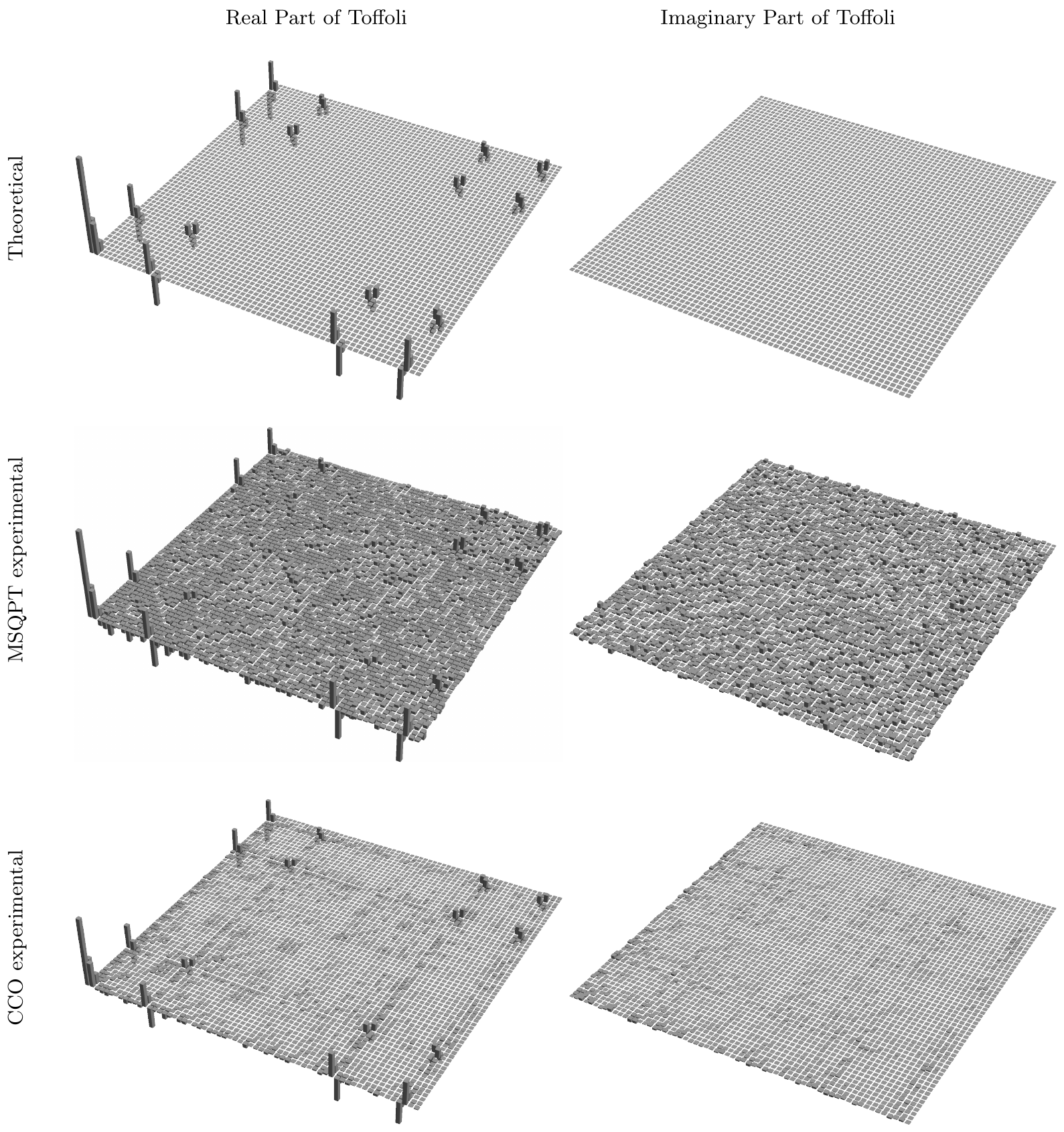} 
\caption{Tomographs corresponding to
the three-qubit
Toffoli gate, with the first and second columns representing 
the real and imaginary part of
the $\chi$ matrix, respectively.
The
first row represents the theoretically constructed $\chi$
matrix while the second and third rows represent the experimentally
constructed $\chi$ matrix obtained by implementing the MSQPT
and the CCO protocols, respectively.
} 
\label{toffoli}
\end{figure*}

Preparation of the five-qubit input state requires finding
the correct decomposition of the unitary operator
$\mathcal{R}_i$ (Fig.~\ref{genckt}) in terms of CNOT gates
and single-qubit rotations, and several decomposition
techniques are
available~\cite{bergholm-pra-2005,shende-ieee-2006}.  We
used the Mathematica package 
{\em UniversalQCompiler}~\cite{plesch-pra-2011,colbeck-pra-2016}  as an
optimization tool to prepare the input state $\vert
\Omega_i \rangle$ from the initial state $\vert 00000
\rangle$.  The quantum circuit to perform
MSQPT of a three-qubit Toffoli gate is given in
Fig.~\ref{ibm4}, corresponding to the five-qubit pure input
state $\vert \Omega_{50} \rangle$ and the observable $E_{15}
= I \otimes \sigma_{z} \otimes \sigma_{y}$.  The system
qubits are denoted by $q[0]$, $q[1]$ and $q[2]$, while the
ancilla qubits are denoted by $q[3]$ and $q[4]$,
respectively.  The first
block in  Fig.~\ref{ibm4} prepares the five-qubit pure input
state $\vert \Omega_{50} \rangle$ while the second block
represents the action of the Toffoli gate on the system
qubits and the last block represents the action of the
quantum map corresponding to the unitary operation $U_{15}=
$CNOT$_{23}$.  A measurement is made on the third qubit in the
$\sigma_{z}$ basis, to compute the quantity ${\rm
Tr}(\sigma_{3z}
U_{15}(\Lambda(\tilde{\rho}_{50})){U_{15}}^{\dagger})$, and
obtain: 

\begin{eqnarray}
{\rm Tr}(E_{15} \Lambda(E_{50} )) &=& 8 
{\rm Tr}(E_{15} \Lambda(\tilde{\rho}_{50})) \nonumber \\
 &=& 8{\rm Tr}(\sigma_{3z} U_{15}(\Lambda(\tilde{\rho}_{50})){U_{15}}^{\dagger})
\end{eqnarray}

All the ${\rm Tr}(E_k \Lambda(E_i))=\langle E^i_k\rangle$
can be computed in a similar fashion, corresponding to the
desired average survival probability $F_{mn}$.  
The list of all unitary operations $U_{i}$, 
corresponding to all quantum maps for the three-qubit system 
can be found in~\cite{singh-pra-2018}.
The
experimentally obtained $ 64\times64$ dimensional $\chi$
matrix corresponding to the three-qubit Toffoli gate is
depicted in Fig.~\ref{toffoli}  as a bar plot, where the
first and second columns represent the real and
imaginary parts of the $\chi$ matrix, respectively.  The
first row denotes the theoretically constructed process
matrix $\chi_{{\rm the}}$, while the second and third rows
represent the experimentally constructed process matrices
$\chi_{{\rm msqpt}}$ and $\chi_{{\rm cco}}$, respectively.
The experimental gate fidelity for $\chi_{{\rm msqpt}}$
turns out to be 0.589, while the much improved experimental
gate fidelity obtained for the case of $\chi_{{\rm cco}}$
turns out be 0.946. To ensure the correctness of the
circuits, we also simulated all the circuits on the IBM
simulator, with a simulation  fidelity of 0.98. 

The total number of readouts to determine a specific
element of the three-qubit
process matrix  with high precision, using the standard
QPT protocol, is $63
\times 63 = 3969 $.  For three qubits, the cardinality of
the set of quantum 2-design states is 72 (9 MUB sets each
having a cardinality of 8).  Determining the real and
imaginary parts of $F_{mn}$ respectively for three qubits
requires a total of 288 state preparations using the SQPT
protocol.  Further to estimate the overlap with original
state $\Phi_j$, 7 readouts need to be performed.  Hence the
total number of readouts required to determine a 
specific element of the three-qubit
process matrix with high precision, using the SQPT
method, is $288 \times 7 =
2016$.  For the MSQPT method, the total number of states we
need to prepare are 63 (corresponding to the complete set of
basis operators) and the total number of readouts required
is 504 (for each MUB set we need to perform 56 readouts, so
the total number of readouts is $9 \times 56 = 504$).  This
makes the MSQPT method vastly more efficient as compared to the standard QPT
and SQPT protocols.  The experimental complexity and number
of ancilla qubits required to determine a specific element 
of the 
process matrix with high precision for
a three-qubit system, using the MSQPT
method, are compared with the
standard QPT and SQPT methods in Table~\ref{complexity3}.

\section{Conclusions}
\label{concl}
We proposed a quantum circuit to efficiently
implement the MSQPT protocol
which reduces the experimental cost of performing 
standard SQPT.
We implemented the MSQPT protocol 
on the IBM quantum processor.  
The system was
prepared in a mixed state corresponding to all Pauli
operators and the MSQPT protocol to perform elementwise
process tomography of two- and three-qubit quantum gates was
successfully implemented.
Our experimental results indicate that MSQPT is
substantially more
efficient as compared to SQPT and standard methods, when
estimating specific elements of the process matrix with
high precision.  We also showed that one can utilize the
full process matrix obtained experimentally via MSQPT, to
solve the $l_2$-norm minimization problem and reconstruct
the underlying true quantum process.  The MSQPT method opens
up several avenues for future applications such as finding
an optimal set of basis operators, developing generalized
algorithms to find all sets of quantum maps to perform
efficient measurements, and finding efficient decompositions
of unitaries using the set of available quantum gates for
easy experimental implementation.
\begin{acknowledgments}
The authors acknowledge support from the IBM quantum experience
team for providing the experimental platform.  
K.S. acknowledges financial support from the
Prime Minister's Research Fellowship (PMRF) scheme of the
Government of India.
Arvind
acknowledges financial support from
DST/ICPS/QuST/Theme-1/2019/General Project number Q-68.
K.D. acknowledges financial support from
DST/ICPS/QuST/Theme-2/2019/General Project number Q-74.
\end{acknowledgments}
%

\begin{widetext}
\appendix
\section{Three-Qubit MSQPT}
\label{sec.appendix1}
\allowdisplaybreaks
{\small
\begin{align*}
& 
 \vert \Omega_1 \rangle =[(0,0,0,\frac{1}{2 \sqrt{2}},0,0,0,\frac{1}{2 \sqrt{2}},0,0,\frac{1}{2 \sqrt{2}},0,0,0,\frac{1}{2 \sqrt{2}},0,0,\frac{1}{2 \sqrt{2}},0,0,0,\frac{1}{2 \sqrt{2}},0,0,\frac{1}{2 \sqrt{2}},0,0,0,\frac{1}{2 \sqrt{2}},0,0,0)]^{T}, \\&
 \vert \Omega_2 \rangle =[(0,0,0,-\frac{i}{2 \sqrt{2}},0,0,0,\frac{1}{2 \sqrt{2}},0,0,-\frac{i}{2 \sqrt{2}},0,0,0,\frac{1}{2 \sqrt{2}},0,0,-\frac{i}{2 \sqrt{2}},0,0,0,\frac{1}{2 \sqrt{2}},0,0,-\frac{i}{2 \sqrt{2}},0,0,0,\frac{1}{2 \sqrt{2}},0,0,0)]^{T},
 \\&
 \vert \Omega_3 \rangle=[(0,0,0,\frac{1}{2},0,0,0,0,0,0,\frac{1}{2},0,0,0,0,0,0,\frac{1}{2},0,0,0,0,0,0,\frac{1}{2},0,0,0,0,0,0,0)]^{T}, 
 \\&
 \vert \Omega_4 \rangle=[(0,0,0,\frac{1}{2 \sqrt{2}},0,0,\frac{1}{2 \sqrt{2}},0,0,0,0,\frac{1}{2 \sqrt{2}},0,0,\frac{1}{2 \sqrt{2}},0,0,\frac{1}{2 \sqrt{2}},0,0,\frac{1}{2 \sqrt{2}},0,0,0,0,\frac{1}{2 \sqrt{2}},0,0,\frac{1}{2 \sqrt{2}},0,0,0)]^{T}, 
 \\&
 \vert \Omega_5 \rangle=[(0,0,\frac{1}{2 \sqrt{2}},0,0,0,0,\frac{1}{2 \sqrt{2}},0,0,0,\frac{1}{2 \sqrt{2}},0,0,\frac{1}{2 \sqrt{2}},0,\frac{1}{2 \sqrt{2}},0,0,0,0,\frac{1}{2 \sqrt{2}},0,0,0,\frac{1}{2 \sqrt{2}},0,0,\frac{1}{2 \sqrt{2}},0,0,0)]^{T}, 
\\&
\vert \Omega_6 \rangle=[(0,0,-\frac{i}{2 \sqrt{2}},0,0,0,0,\frac{i}{2 \sqrt{2}},0,0,0,\frac{1}{2 \sqrt{2}},0,0,\frac{1}{2 \sqrt{2}},0,-\frac{i}{2 \sqrt{2}},0,0,0,0,\frac{i}{2 \sqrt{2}},0,0,0,\frac{1}{2 \sqrt{2}},0,0,\frac{1}{2 \sqrt{2}},0,0,0)]^{T}, 
\\&
\vert \Omega_7 \rangle=[(0,0,0,\frac{1}{2 \sqrt{2}},0,0,-\frac{1}{2 \sqrt{2}},0,0,0,0,\frac{1}{2 \sqrt{2}},0,0,\frac{1}{2 \sqrt{2}},0,0,\frac{1}{2 \sqrt{2}},0,0,-\frac{1}{2 \sqrt{2}},0,0,0,0,\frac{1}{2 \sqrt{2}},0,0,\frac{1}{2 \sqrt{2}},0,0,0)]^{T}, 
\\&
\vert \Omega_8 \rangle=[(0,0,0,-\frac{i}{2 \sqrt{2}},0,0,-\frac{i}{2 \sqrt{2}},0,0,0,0,\frac{1}{2 \sqrt{2}},0,0,\frac{1}{2 \sqrt{2}},0,0,-\frac{i}{2 \sqrt{2}},0,0,-\frac{i}{2 \sqrt{2}},0,0,0,0,\frac{1}{2 \sqrt{2}},0,0,\frac{1}{2 \sqrt{2}},0,0,0)]^{T}, 
\\&
\vert \Omega_9 \rangle=[(0,0,-\frac{i}{2 \sqrt{2}},0,0,0,0,-\frac{i}{2 \sqrt{2}},0,0,0,\frac{1}{2 \sqrt{2}},0,0,\frac{1}{2 \sqrt{2}},0,-\frac{i}{2 \sqrt{2}},0,0,0,0,-\frac{i}{2 \sqrt{2}},0,0,0,\frac{1}{2 \sqrt{2}},0,0,\frac{1}{2 \sqrt{2}},0,0,0)]^{T}, 
\\&
\vert \Omega_{10} \rangle=[(0,0,-\frac{1}{2 \sqrt{2}},0,0,0,0,\frac{1}{2 \sqrt{2}},0,0,0,\frac{1}{2 \sqrt{2}},0,0,\frac{1}{2 \sqrt{2}},0,-\frac{1}{2 \sqrt{2}},0,0,0,0,\frac{1}{2 \sqrt{2}},0,0,0,\frac{1}{2 \sqrt{2}},0,0,\frac{1}{2 \sqrt{2}},0,0,0)]^{T}, 
\\&
\vert \Omega_{11} \rangle=[(0,0,0,-\frac{i}{2 \sqrt{2}},0,0,\frac{i}{2 \sqrt{2}},0,0,0,0,\frac{1}{2 \sqrt{2}},0,0,\frac{1}{2 \sqrt{2}},0,0,-\frac{i}{2 \sqrt{2}},0,0,\frac{i}{2 \sqrt{2}},0,0,0,0,\frac{1}{2 \sqrt{2}},0,0,\frac{1}{2 \sqrt{2}},0,0,0)]^{T}, 
\\&
\vert \Omega_{12} \rangle=[(0,0,0,\frac{1}{2},0,0,\frac{1}{2},0,0,0,0,0,0,0,0,0,0,\frac{1}{2},0,0,\frac{1}{2},0,0,0,0,0,0,0,0,0,0,0)]^{T}, 
\\&
\vert \Omega_{13} \rangle=[(0,0,0,\frac{1}{2 \sqrt{2}},0,0,0,\frac{1}{2 \sqrt{2}},0,0,-\frac{1}{2 \sqrt{2}},0,0,0,\frac{1}{2 \sqrt{2}},0,0,\frac{1}{2 \sqrt{2}},0,0,0,\frac{1}{2 \sqrt{2}},0,0,-\frac{1}{2 \sqrt{2}},0,0,0,\frac{1}{2 \sqrt{2}},0,0,0)]^{T},
\\&
\vert \Omega_{14} \rangle=[(0,0,0,\frac{1}{2 \sqrt{2}},0,0,0,\frac{1}{2 \sqrt{2}},0,0,-\frac{1}{2 \sqrt{2}},0,0,0,\frac{1}{2 \sqrt{2}},0,0,\frac{1}{2 \sqrt{2}},0,0,0,\frac{1}{2 \sqrt{2}},0,0,-\frac{1}{2 \sqrt{2}},0,0,0,\frac{1}{2 \sqrt{2}},0,0,0)]^{T}, 
\\&
\vert \Omega_{15} \rangle=[(0,0,0,\frac{1}{2},0,0,0,0,0,0,0,0,0,0,\frac{1}{2},0,0,\frac{1}{2},0,0,0,0,0,0,0,0,0,0,\frac{1}{2},0,0,0)]^{T}, 
\\&
\vert \Omega_{16} \rangle=[(0,0,0,\frac{1}{2 \sqrt{2}},0,0,\frac{1}{2 \sqrt{2}},0,0,\frac{1}{2 \sqrt{2}},0,0,\frac{1}{2 \sqrt{2}},0,0,0,0,0,0,\frac{1}{2 \sqrt{2}},0,0,\frac{1}{2 \sqrt{2}},0,0,\frac{1}{2 \sqrt{2}},0,0,\frac{1}{2 \sqrt{2}},0,0,0)]^{T}, 
\\&
\vert \Omega_{17} \rangle=[(0,0,\frac{1}{2 \sqrt{2}},0,0,0,0,\frac{1}{2 \sqrt{2}},\frac{1}{2 \sqrt{2}},0,0,0,0,\frac{1}{2 \sqrt{2}},0,0,0,0,0,\frac{1}{2 \sqrt{2}},0,0,\frac{1}{2 \sqrt{2}},0,0,\frac{1}{2 \sqrt{2}},0,0,\frac{1}{2 \sqrt{2}},0,0,0)]^{T}, 
\\&
\vert \Omega_{18} \rangle=[(0,0,-\frac{i}{2 \sqrt{2}},0,0,0,0,\frac{i}{2 \sqrt{2}},-\frac{i}{2 \sqrt{2}},0,0,0,0,\frac{i}{2 \sqrt{2}},0,0,0,0,0,\frac{1}{2 \sqrt{2}},0,0,\frac{1}{2 \sqrt{2}},0,0,\frac{1}{2 \sqrt{2}},0,0,\frac{1}{2 \sqrt{2}},0,0,0)]^{T}, 
\\&
\vert \Omega_{19} \rangle =[(0,0,0,\frac{1}{2 \sqrt{2}},0,0,-\frac{1}{2 \sqrt{2}},0,0,\frac{1}{2 \sqrt{2}},0,0,-\frac{1}{2 \sqrt{2}},0,0,0,0,0,0,\frac{1}{2 \sqrt{2}},0,0,\frac{1}{2 \sqrt{2}},0,0,\frac{1}{2 \sqrt{2}},0,0,\frac{1}{2 \sqrt{2}},0,0,0)]^{T}, 
\\&
\vert \Omega_{20} \rangle=[(0,\frac{1}{2 \sqrt{2}},0,0,\frac{1}{2 \sqrt{2}},0,0,0,0,0,0,\frac{1}{2 \sqrt{2}},0,0,\frac{1}{2 \sqrt{2}},0,0,0,0,\frac{1}{2 \sqrt{2}},0,0,\frac{1}{2 \sqrt{2}},0,0,\frac{1}{2 \sqrt{2}},0,0,\frac{1}{2 \sqrt{2}},0,0,0)]^{T}, 
\\&
\vert \Omega_{21} \rangle=[(\frac{1}{2 \sqrt{2}},0,0,0,0,\frac{1}{2 \sqrt{2}},0,0,0,0,\frac{1}{2 \sqrt{2}},0,0,0,0,\frac{1}{2 \sqrt{2}},0,0,0,\frac{1}{2 \sqrt{2}},0,0,\frac{1}{2 \sqrt{2}},0,0,\frac{1}{2 \sqrt{2}},0,0,\frac{1}{2 \sqrt{2}},0,0,0)]^{T}, 
\\&
\vert \Omega_{22} \rangle=[(-\frac{i}{2 \sqrt{2}},0,0,0,0,\frac{i}{2 \sqrt{2}},0,0,0,0,-\frac{i}{2 \sqrt{2}},0,0,0,0,\frac{i}{2 \sqrt{2}},0,0,0,\frac{1}{2 \sqrt{2}},0,0,\frac{1}{2 \sqrt{2}},0,0,\frac{1}{2 \sqrt{2}},0,0,\frac{1}{2 \sqrt{2}},0,0,0)]^{T}, 
\\&
\vert \Omega_{23} \rangle=[(0,\frac{1}{2 \sqrt{2}},0,0,-\frac{1}{2 \sqrt{2}},0,0,0,0,0,0,\frac{1}{2 \sqrt{2}},0,0,-\frac{1}{2 \sqrt{2}},0,0,0,0,\frac{1}{2 \sqrt{2}},0,0,\frac{1}{2 \sqrt{2}},0,0,\frac{1}{2 \sqrt{2}},0,0,\frac{1}{2 \sqrt{2}},0,0,0)]^{T}, 
\\&
\vert \Omega_{24} \rangle=[(0,-\frac{i}{2 \sqrt{2}},0,0,-\frac{i}{2 \sqrt{2}},0,0,0,0,0,0,\frac{i}{2 \sqrt{2}},0,0,\frac{i}{2 \sqrt{2}},0,0,0,0,\frac{1}{2 \sqrt{2}},0,0,\frac{1}{2 \sqrt{2}},0,0,\frac{1}{2 \sqrt{2}},0,0,\frac{1}{2 \sqrt{2}},0,0,0)]^{T}, 
\\&
\vert \Omega_{25} \rangle=[(-\frac{i}{2 \sqrt{2}},0,0,0,0,-\frac{i}{2 \sqrt{2}},0,0,0,0,\frac{i}{2 \sqrt{2}},0,0,0,0,\frac{i}{2 \sqrt{2}},0,0,0,\frac{1}{2 \sqrt{2}},0,0,\frac{1}{2 \sqrt{2}},0,0,\frac{1}{2 \sqrt{2}},0,0,\frac{1}{2 \sqrt{2}},0,0,0)]^{T}, 
\\&
\vert \Omega_{26} \rangle=[(-\frac{1}{2 \sqrt{2}},0,0,0,0,\frac{1}{2 \sqrt{2}},0,0,0,0,\frac{1}{2 \sqrt{2}},0,0,0,0,-\frac{1}{2 \sqrt{2}},0,0,0,\frac{1}{2 \sqrt{2}},0,0,\frac{1}{2 \sqrt{2}},0,0,\frac{1}{2 \sqrt{2}},0,0,\frac{1}{2 \sqrt{2}},0,0,0)]^{T}, 
\\&
\vert \Omega_{27} \rangle=[(0,-\frac{i}{2 \sqrt{2}},0,0,\frac{i}{2 \sqrt{2}},0,0,0,0,0,0,\frac{i}{2 \sqrt{2}},0,0,-\frac{i}{2 \sqrt{2}},0,0,0,0,\frac{1}{2 \sqrt{2}},0,0,\frac{1}{2 \sqrt{2}},0,0,\frac{1}{2 \sqrt{2}},0,0,\frac{1}{2 \sqrt{2}},0,0,0)]^{T}, 
\\&
\vert \Omega_{28} \rangle=[(0,0,0,\frac{1}{2 \sqrt{2}},0,0,\frac{1}{2 \sqrt{2}},0,0,-\frac{1}{2 \sqrt{2}},0,0,-\frac{1}{2 \sqrt{2}},0,0,0,0,0,0,\frac{1}{2 \sqrt{2}},0,0,\frac{1}{2 \sqrt{2}},0,0,\frac{1}{2 \sqrt{2}},0,0,\frac{1}{2 \sqrt{2}},0,0,0)]^{T}, 
\\&
\vert \Omega_{29} \rangle=[(0,0,\frac{1}{2 \sqrt{2}},0,0,0,0,\frac{1}{2 \sqrt{2}},-\frac{1}{2 \sqrt{2}},0,0,0,0,-\frac{1}{2 \sqrt{2}},0,0,0,0,0,\frac{1}{2 \sqrt{2}},0,0,\frac{1}{2 \sqrt{2}},0,0,\frac{1}{2 \sqrt{2}},0,0,\frac{1}{2 \sqrt{2}},0,0,0)]^{T}, 
\\&
\vert \Omega_{30} \rangle=[(0,0,-\frac{i}{2 \sqrt{2}},0,0,0,0,\frac{i}{2 \sqrt{2}},\frac{i}{2 \sqrt{2}},0,0,0,0,-\frac{i}{2 \sqrt{2}},0,0,0,0,0,\frac{1}{2 \sqrt{2}},0,0,\frac{1}{2 \sqrt{2}},0,0,\frac{1}{2 \sqrt{2}},0,0,\frac{1}{2 \sqrt{2}},0,0,0)]^{T}, 
\\&
\vert \Omega_{31} \rangle=[(0,0,0,\frac{1}{2 \sqrt{2}},0,0,-\frac{1}{2 \sqrt{2}},0,0,-\frac{1}{2 \sqrt{2}},0,0,\frac{1}{2 \sqrt{2}},0,0,0,0,0,0,\frac{1}{2 \sqrt{2}},0,0,\frac{1}{2 \sqrt{2}},0,0,\frac{1}{2 \sqrt{2}},0,0,\frac{1}{2 \sqrt{2}},0,0,0)]^{T}, 
\\&
\vert \Omega_{32} \rangle=[(0,0,0,-\frac{i}{2 \sqrt{2}},0,0,-\frac{i}{2 \sqrt{2}},0,0,-\frac{i}{2 \sqrt{2}},0,0,-\frac{i}{2 \sqrt{2}},0,0,0,0,0,0,\frac{1}{2 \sqrt{2}},0,0,\frac{1}{2 \sqrt{2}},0,0,\frac{1}{2 \sqrt{2}},0,0,\frac{1}{2 \sqrt{2}},0,0,0)]^{T}, 
\\&
\vert \Omega_{33} \rangle=[(0,0,-\frac{i}{2 \sqrt{2}},0,0,0,0,-\frac{i}{2 \sqrt{2}},-\frac{i}{2 \sqrt{2}},0,0,0,0,-\frac{i}{2 \sqrt{2}},0,0,0,0,0,\frac{1}{2 \sqrt{2}},0,0,\frac{1}{2 \sqrt{2}},0,0,\frac{1}{2 \sqrt{2}},0,0,\frac{1}{2 \sqrt{2}},0,0,0)]^{T}, 
\\&
\vert \Omega_{34} \rangle=[(0,0,-\frac{1}{2 \sqrt{2}},0,0,0,0,\frac{1}{2 \sqrt{2}},-\frac{1}{2 \sqrt{2}},0,0,0,0,\frac{1}{2 \sqrt{2}},0,0,0,0,0,\frac{1}{2 \sqrt{2}},0,0,\frac{1}{2 \sqrt{2}},0,0,\frac{1}{2 \sqrt{2}},0,0,\frac{1}{2 \sqrt{2}},0,0,0)]^{T}, 
\\&
\vert \Omega_{35} \rangle=[(0,0,0,-\frac{i}{2 \sqrt{2}},0,0,\frac{i}{2 \sqrt{2}},0,0,-\frac{i}{2 \sqrt{2}},0,0,\frac{i}{2 \sqrt{2}},0,0,0,0,0,0,\frac{1}{2 \sqrt{2}},0,0,\frac{1}{2 \sqrt{2}},0,0,\frac{1}{2 \sqrt{2}},0,0,\frac{1}{2 \sqrt{2}},0,0,0)]^{T}, 
\\&
\vert \Omega_{36} \rangle=[(0,-\frac{i}{2 \sqrt{2}},0,0,-\frac{i}{2 \sqrt{2}},0,0,0,0,0,0,-\frac{i}{2 \sqrt{2}},0,0,-\frac{i}{2 \sqrt{2}},0,0,0,0,\frac{1}{2 \sqrt{2}},0,0,\frac{1}{2 \sqrt{2}},0,0,\frac{1}{2 \sqrt{2}},0,0,\frac{1}{2 \sqrt{2}},0,0,0)]^{T}, 
\\&
\vert \Omega_{37} \rangle=[(-\frac{i}{2 \sqrt{2}},0,0,0,0,-\frac{i}{2 \sqrt{2}},0,0,0,0,-\frac{i}{2 \sqrt{2}},0,0,0,0,-\frac{i}{2 \sqrt{2}},0,0,0,\frac{1}{2 \sqrt{2}},0,0,\frac{1}{2 \sqrt{2}},0,0,\frac{1}{2 \sqrt{2}},0,0,\frac{1}{2 \sqrt{2}},0,0,0)]^{T}, 
\\&
\vert \Omega_{38} \rangle=[(-\frac{1}{2 \sqrt{2}},0,0,0,0,\frac{1}{2 \sqrt{2}},0,0,0,0,-\frac{1}{2 \sqrt{2}},0,0,0,0,\frac{1}{2 \sqrt{2}},0,0,0,\frac{1}{2 \sqrt{2}},0,0,\frac{1}{2 \sqrt{2}},0,0,\frac{1}{2 \sqrt{2}},0,0,\frac{1}{2 \sqrt{2}},0,0,0)]^{T}, 
\\&
\vert \Omega_{39} \rangle=[(0,-\frac{i}{2 \sqrt{2}},0,0,\frac{i}{2 \sqrt{2}},0,0,0,0,0,0,-\frac{i}{2 \sqrt{2}},0,0,\frac{i}{2 \sqrt{2}},0,0,0,0,\frac{1}{2 \sqrt{2}},0,0,\frac{1}{2 \sqrt{2}},0,0,\frac{1}{2 \sqrt{2}},0,0,\frac{1}{2 \sqrt{2}},0,0,0)]^{T}, 
\\&
\vert \Omega_{40} \rangle=[(0,-\frac{1}{2 \sqrt{2}},0,0,-\frac{1}{2 \sqrt{2}},0,0,0,0,0,0,\frac{1}{2 \sqrt{2}},0,0,\frac{1}{2 \sqrt{2}},0,0,0,0,\frac{1}{2 \sqrt{2}},0,0,\frac{1}{2 \sqrt{2}},0,0,\frac{1}{2 \sqrt{2}},0,0,\frac{1}{2 \sqrt{2}},0,0,0)]^{T}, 
\\&
\vert \Omega_{41} \rangle=[(-\frac{1}{2 \sqrt{2}},0,0,0,0,-\frac{1}{2 \sqrt{2}},0,0,0,0,\frac{1}{2 \sqrt{2}},0,0,0,0,\frac{1}{2 \sqrt{2}},0,0,0,\frac{1}{2 \sqrt{2}},0,0,\frac{1}{2 \sqrt{2}},0,0,\frac{1}{2 \sqrt{2}},0,0,\frac{1}{2 \sqrt{2}},0,0,0)]^{T}, 
\\&
\vert \Omega_{42} \rangle=[(\frac{i}{2 \sqrt{2}},0,0,0,0,-\frac{i}{2 \sqrt{2}},0,0,0,0,-\frac{i}{2 \sqrt{2}},0,0,0,0,\frac{i}{2 \sqrt{2}},0,0,0,\frac{1}{2 \sqrt{2}},0,0,\frac{1}{2 \sqrt{2}},0,0,\frac{1}{2 \sqrt{2}},0,0,\frac{1}{2 \sqrt{2}},0,0,0)]^{T}, 
\\&
\vert \Omega_{43} \rangle=[(0,-\frac{1}{2 \sqrt{2}},0,0,\frac{1}{2 \sqrt{2}},0,0,0,0,0,0,\frac{1}{2 \sqrt{2}},0,0,-\frac{1}{2 \sqrt{2}},0,0,0,0,\frac{1}{2 \sqrt{2}},0,0,\frac{1}{2 \sqrt{2}},0,0,\frac{1}{2 \sqrt{2}},0,0,\frac{1}{2 \sqrt{2}},0,0,0)]^{T}, 
\\&
\vert \Omega_{44} \rangle=[(0,0,0,-\frac{i}{2 \sqrt{2}},0,0,-\frac{i}{2 \sqrt{2}},0,0,\frac{i}{2 \sqrt{2}},0,0,\frac{i}{2 \sqrt{2}},0,0,0,0,0,0,\frac{1}{2 \sqrt{2}},0,0,\frac{1}{2 \sqrt{2}},0,0,\frac{1}{2 \sqrt{2}},0,0,\frac{1}{2 \sqrt{2}},0,0,0)]^{T}, 
\\&
\vert \Omega_{45} \rangle=[(0,0,-\frac{i}{2 \sqrt{2}},0,0,0,0,-\frac{i}{2 \sqrt{2}},\frac{i}{2 \sqrt{2}},0,0,0,0,\frac{i}{2 \sqrt{2}},0,0,0,0,0,\frac{1}{2 \sqrt{2}},0,0,\frac{1}{2 \sqrt{2}},0,0,\frac{1}{2 \sqrt{2}},0,0,\frac{1}{2 \sqrt{2}},0,0,0)]^{T}, 
\\&
\vert \Omega_{46} \rangle=[(0,0,-\frac{1}{2 \sqrt{2}},0,0,0,0,\frac{1}{2 \sqrt{2}},\frac{1}{2 \sqrt{2}},0,0,0,0,-\frac{1}{2 \sqrt{2}},0,0,0,0,0,\frac{1}{2 \sqrt{2}},0,0,\frac{1}{2 \sqrt{2}},0,0,\frac{1}{2 \sqrt{2}},0,0,\frac{1}{2 \sqrt{2}},0,0,0)]^{T}, 
\\&
\vert \Omega_{47} \rangle=[(0,0,0,-\frac{i}{2 \sqrt{2}},0,0,\frac{i}{2 \sqrt{2}},0,0,\frac{i}{2 \sqrt{2}},0,0,-\frac{i}{2 \sqrt{2}},0,0,0,0,0,0,\frac{1}{2 \sqrt{2}},0,0,\frac{1}{2 \sqrt{2}},0,0,\frac{1}{2 \sqrt{2}},0,0,\frac{1}{2 \sqrt{2}},0,0,0)]^{T},
\\&
\vert \Omega_{48} \rangle=[(0,0,0,\frac{1}{2},0,0,\frac{1}{2},0,0,\frac{1}{2},0,0,\frac{1}{2},0,0,0,0,0,0,0,0,0,0,0,0,0,0,0,0,0,0,0)]^{T}, 
\\&
\vert \Omega_{49} \rangle=[(0,0,0,\frac{1}{2 \sqrt{2}},0,0,0,\frac{1}{2 \sqrt{2}},0,0,\frac{1}{2 \sqrt{2}},0,0,0,\frac{1}{2 \sqrt{2}},0,0,-\frac{1}{2 \sqrt{2}},0,0,0,\frac{1}{2 \sqrt{2}},0,0,-\frac{1}{2 \sqrt{2}},0,0,0,\frac{1}{2 \sqrt{2}},0,0,0)]^{T}, 
\\&
\vert \Omega_{50} \rangle=[(0,0,0,-\frac{i}{2 \sqrt{2}},0,0,0,\frac{1}{2 \sqrt{2}},0,0,-\frac{i}{2 \sqrt{2}},0,0,0,\frac{1}{2 \sqrt{2}},0,0,\frac{i}{2 \sqrt{2}},0,0,0,\frac{1}{2 \sqrt{2}},0,0,\frac{i}{2 \sqrt{2}},0,0,0,\frac{1}{2 \sqrt{2}},0,0,0)]^{T}, 
\\&
\vert \Omega_{51} \rangle=[(0,0,0,\frac{1}{2},0,0,0,0,0,0,\frac{1}{2},0,0,0,0,0,0,0,0,0,0,\frac{1}{2},0,0,0,0,0,0,\frac{1}{2},0,0,0)]^{T}, 
\\&
\vert \Omega_{52} \rangle=[(0,0,0,\frac{1}{2 \sqrt{2}},0,0,\frac{1}{2 \sqrt{2}},0,0,0,0,\frac{1}{2 \sqrt{2}},0,0,\frac{1}{2 \sqrt{2}},0,0,-\frac{1}{2 \sqrt{2}},0,0,-\frac{1}{2 \sqrt{2}},0,0,0,0,\frac{1}{2 \sqrt{2}},0,0,\frac{1}{2 \sqrt{2}},0,0,0)]^{T}, 
\\&
\vert \Omega_{53} \rangle=[(0,0,\frac{1}{2 \sqrt{2}},0,0,0,0,\frac{1}{2 \sqrt{2}},0,0,0,\frac{1}{2 \sqrt{2}},0,0,\frac{1}{2 \sqrt{2}},0,-\frac{1}{2 \sqrt{2}},0,0,0,0,-\frac{1}{2 \sqrt{2}},0,0,0,\frac{1}{2 \sqrt{2}},0,0,\frac{1}{2 \sqrt{2}},0,0,0)]^{T}, 
\\&
\vert \Omega_{54} \rangle=[(0,0,-\frac{i}{2 \sqrt{2}},0,0,0,0,\frac{i}{2 \sqrt{2}},0,0,0,\frac{1}{2 \sqrt{2}},0,0,\frac{1}{2 \sqrt{2}},0,\frac{i}{2 \sqrt{2}},0,0,0,0,-\frac{i}{2 \sqrt{2}},0,0,0,\frac{1}{2 \sqrt{2}},0,0,\frac{1}{2 \sqrt{2}},0,0,0)]^{T}, 
\\&
\vert \Omega_{55} \rangle=[(0,0,0,\frac{1}{2 \sqrt{2}},0,0,-\frac{1}{2 \sqrt{2}},0,0,0,0,\frac{1}{2 \sqrt{2}},0,0,\frac{1}{2 \sqrt{2}},0,0,-\frac{1}{2 \sqrt{2}},0,0,\frac{1}{2 \sqrt{2}},0,0,0,0,\frac{1}{2 \sqrt{2}},0,0,\frac{1}{2 \sqrt{2}},0,0,0)]^{T}, 
\\&
\vert \Omega_{56} \rangle=[(0,0,0,-\frac{i}{2 \sqrt{2}},0,0,-\frac{i}{2 \sqrt{2}},0,0,0,0,\frac{1}{2 \sqrt{2}},0,0,\frac{1}{2 \sqrt{2}},0,0,\frac{i}{2 \sqrt{2}},0,0,\frac{i}{2 \sqrt{2}},0,0,0,0,\frac{1}{2 \sqrt{2}},0,0,\frac{1}{2 \sqrt{2}},0,0,0)]^{T}, 
\\&
\vert \Omega_{57} \rangle=[(0,0,-\frac{i}{2 \sqrt{2}},0,0,0,0,-\frac{i}{2 \sqrt{2}},0,0,0,\frac{1}{2 \sqrt{2}},0,0,\frac{1}{2 \sqrt{2}},0,\frac{i}{2 \sqrt{2}},0,0,0,0,\frac{i}{2 \sqrt{2}},0,0,0,\frac{1}{2 \sqrt{2}},0,0,\frac{1}{2 \sqrt{2}},0,0,0)]^{T}, 
\\&
\vert \Omega_{58} \rangle=[(0,0,-\frac{1}{2 \sqrt{2}},0,0,0,0,\frac{1}{2 \sqrt{2}},0,0,0,\frac{1}{2 \sqrt{2}},0,0,\frac{1}{2 \sqrt{2}},0,\frac{1}{2 \sqrt{2}},0,0,0,0,-\frac{1}{2 \sqrt{2}},0,0,0,\frac{1}{2 \sqrt{2}},0,0,\frac{1}{2 \sqrt{2}},0,0,0)]^{T}, 
\\&
\vert \Omega_{59} \rangle=[(0,0,0,-\frac{i}{2 \sqrt{2}},0,0,\frac{i}{2 \sqrt{2}},0,0,0,0,\frac{1}{2 \sqrt{2}},0,0,\frac{1}{2 \sqrt{2}},0,0,\frac{i}{2 \sqrt{2}},0,0,-\frac{i}{2 \sqrt{2}},0,0,0,0,\frac{1}{2 \sqrt{2}},0,0,\frac{1}{2 \sqrt{2}},0,0,0)]^{T}, 
\\&
\vert \Omega_{60} \rangle=[(0,0,0,\frac{1}{2},0,0,\frac{1}{2},0,0,0,0,0,0,0,0,0,0,0,0,0,0,0,0,0,0,\frac{1}{2},0,0,\frac{1}{2},0,0,0)]^{T}, 
\\&
\vert \Omega_{61} \rangle=[(0,0,0,\frac{1}{2 \sqrt{2}},0,0,0,\frac{1}{2 \sqrt{2}},0,0,-\frac{1}{2 \sqrt{2}},0,0,0,\frac{1}{2 \sqrt{2}},0,0,-\frac{1}{2 \sqrt{2}},0,0,0,\frac{1}{2 \sqrt{2}},0,0,\frac{1}{2 \sqrt{2}},0,0,0,\frac{1}{2 \sqrt{2}},0,0,0)]^{T}, 
\\&
\vert \Omega_{52} \rangle=[(0,0,0,-\frac{i}{2 \sqrt{2}},0,0,0,\frac{1}{2 \sqrt{2}},0,0,\frac{i}{2 \sqrt{2}},0,0,0,\frac{1}{2 \sqrt{2}},0,0,\frac{i}{2 \sqrt{2}},0,0,0,\frac{1}{2 \sqrt{2}},0,0,-\frac{i}{2 \sqrt{2}},0,0,0,\frac{1}{2 \sqrt{2}},0,0,0)]^{T}, 
\\&
\vert \Omega_{63} \rangle=[(0,0,0,\frac{1}{2},0,0,0,0,0,0,0,0,0,0,\frac{1}{2},0,0,0,0,0,0,\frac{1}{2},0,0,\frac{1}{2},0,0,0,0,0,0,0)]^{T}, 
\end{align*}
}

\end{widetext}

\end{document}